\documentclass[twoside,twocolumn]{article}
\usepackage{times}
\usepackage{latexsym}
\usepackage{amsbsy}
\usepackage{ifthen}
\usepackage{archiv2e}

\usepackage[dvips]{epsfig}

\begin{document}
%
\Eingangsdatum{Month 00, 2004}  
\Kolumnentitel{
A. Lakhtakia, T.G. Mackay:
Size--dependent Bruggeman formula}
\Autoren{\Name{f.}{vv}{ll}{j.}} \Sachgebiet{1} \Band{58}
\Jahr{2004} \Heft{1} \Ersteseite{1}
%
\Letzteseite{3}
%
%


\def\##1{\underline #1}
\def\=#1{\underline{\underline #1}}

\def\eps{\epsilon}
\def\epso{\epsilon_0}
\def\muo{\mu_0}
\def\ko{k_0}
\def\kosq{k_0^2}
\def\lambdao{\lambda_0}
\def\etao{\eta_0}
\def\.{\mbox{ \tiny{$^\bullet$} }}

\def\curl{\nabla\times}
\def\div{\nabla \mbox{ \tiny{$^\bullet$} }}

\def\ux{\#{u}_x}
\def\uy{\#{u}_y}
\def\uz{\#{u}_z}
\def\up{\#{u}_+}
\def\um{\#{u}_-}

\def\le{\left(}
\def\ri{\right)}
\def\les{\left[}
\def\ris{\right]}
\def\lec{\left\{}
\def\ric{\right\}}

\def\c#1{\cite{#1}}
\def\l#1{\label{#1}}
\def\r#1{(\ref{#1})}

\newcommand{\mat}[1] {\left[\begin{array}{cccc}
             #1 \end{array}\right]}

\newcommand{\mb}[1]{\mbox{\boldmath$\bf#1$}}



\Keywords{Bruggeman approach, Dielectric--magnetic material, 
Homogenization, Maxwell Garnett
approach, Particulate composite material,
Size dependence}

\Title{Size--Dependent Bruggeman Approach for Dielectric--Magnetic 
Composite Materials}

\Authors{Akhlesh Lakhtakia, Tom G. Mackay}

\Addresses{A. Lakhtakia,
  CATMAS---Computational \& Theoretical
Materials Science Group, Department of Engineering Science \&
Mechanics, Pennsylvania State University, University Park, PA
16802--6812, USA.\\ E--mail: akhlesh@psu.edu\\
T.G. Mackay, School of Mathematics,
University of Edinburgh, \\ Edinburgh EH9 3JZ, United Kingdom\\
  E--mail: T.Mackay@ed.ac.uk\\
Correspondence to Mackay
}

\Abstract{Expressions arising from the Bruggeman approach for the 
homogenization of
dielectric--magnetic composite materials, without ignoring the sizes of the
spherical particles, are presented. These expressions exhibit the 
proper limit behavior. The incorporation
of size dependence is directly responsible for the emergence of 
dielectric--magnetic coupling in the
estimated relative permittivity and permeability of the homogenized 
composite material.}


\Paper


\section{Introduction}
The objective of this communication is to introduce a size--dependent 
variant of the
celebrated Bruggeman approach \cite[Eq. 32]{Br} and thereby couple 
the dielectric and magnetic properties
of a particulate composite material (PCM) with isotropic 
dielectric--magnetic constituent materials.

Homogenization of PCMs has been a continuing theme in 
electromagnetism for about two
centuries \c{LOCM}. The most popular approaches consider the particles
to be vanishingly small, point--like entities
  \c{Neela,Michel00}.  Much of the literature is devoted to dielectric 
PCMs \c{Neela, Ward},
with application to magnetic PCMs following as a result of 
electromagnetic duality \cite[Sec. 4-2.3]{LBel}. When PCMs
with both dielectric and magnetic properties are considered, no 
coupling arises between the two
types of constitutive properties if the particles are vanishingly small.
It is this coupling that has gained importance in the
last few years, with the emergence of metamaterials \c{Walser03}.

Investigation of scattering literature quickly reveals that 
dielectric--magnetic coupling in PCMs emerges only
when particles are explicitly taken to be of nonzero  size 
\c{Grimes1,Grimes2, LakhShan}, although the
particle size must still be
electrically small for  the concept of homogenization to remain valid 
\cite[p. xiii]{LOCM}, \c{Lorenz1875}.
To the best of our knowledge, available
homogenization formulas for dielectric--magnetic PCMs that also account for
dielectric--magnetic coupling are applicable only to dilute composites because
they are set up using the Mossotti--Clausius approach (also called 
the Lorenz--Lorentz approach
and the Maxwell Garnett approach \c{PLB}). Use of the Bruggeman approach is
preferred, while maintaining the particle size as nonzero, for 
nondilute composites \c{PLB}.

Accordingly, in Section \ref{theo} we apply the Bruggeman approach to 
derive  size--dependent homogenization
formulas for dielectric--magnetic PCMs comprising spherical 
particles. Sample results are discussed and conclusions are drawn
therefrom in Section \ref{rescon}. An $\exp(-i\omega t)$ 
time--dependence is implicit, with $\omega$ being
the angular frequency. The free--space (i.e., vacuum)
wavenumber is denoted by $k_0$.

\section{Theory}\label{theo}

Let us consider a particulate composite material with $L$ constituent 
materials. The relative permittivity
and the relative permeability of the $\ell$th constituent material, 
$\ell\in[1,\,L]$, are denoted respectively by $\eps_\ell$ and
$\mu_\ell$, the radius of the spherical particles of that material is 
denoted by $R_\ell$, and the volumetric fraction by
$f_\ell$. Clearly,
\begin{equation}
\sum_{\ell=1}^L \,f_\ell= 1\,.
\end{equation}
  Our task is to estimate $\eps_{HCM}$ and $\mu_{HCM}$, which are
the relative permittivity and the relative permeability of the 
homogenized composite material (HCM).

According to the Bruggeman approach \c{Michel00,LakhShan}, the 
following two equations have to be solved:
\begin{equation}
\sum_{\ell=1}^L f_\ell\,\alpha^{\ell/Br}_e = 0\,,\quad
\sum_{\ell=1}^L f_\ell\, \alpha^{\ell/Br}_h = 0\,.
\label{eqBrugg}
\end{equation}
Here,
$\alpha_e^{a/b}$ and $\alpha_h^{a/b}$ are the polarizability density
and the magnetizability density, respectively, of an electrically 
small sphere of material $a$ embedded in material $b$.
In the limit of the particulate radius tending to zero, expressions 
of these two densities are available as follows \c{LVV90}:
\begin{equation}
\left.\begin{array}{ll}
\alpha_e^{a/b} = 3\eps_b\frac{\eps_a-\eps_b}{\eps_a+2\eps_b}\\[5pt]
\alpha_h^{a/b}=3\mu_b\frac{\mu_a-\mu_b}{\mu_a+2\mu_b}
\end{array}
\right\}\,.
\label{eqpol1}
\end{equation}
However, when the sphere radius is nonzero, the foregoing
expressions mutate to include both the radius $R_a$ of the
embedded sphere and the refractive index
\begin{equation}
n_b=\sqrt{\eps_b\mu_b}
\end{equation}
  of the embedding material; thus \c{LakhShan}
\begin{equation}
\left.\begin{array}{ll}
\alpha_e^{a/b} = 3\eps_b\frac{\eps_a-\eps_b}{\eps_a 
(1-2\tau_{a/b})+2\eps_b(1+\tau_{a/b})}\\[5pt]
\alpha_h^{a/b}=3\mu_b\frac{\mu_a-\mu_b}{\mu_a(1-2\tau_{a/b})+2\mu_b(1+\tau_{a/b})}
\end{array}
\right\}\,,
\label{eqpol2}
\end{equation}
where
\begin{equation}
\tau_{a/b} = (1-ik_0R_an_b) \exp(ik_0R_an_b)-1\,.
\end{equation}
More complicated expressions than \r{eqpol2} can be devised by using 
the Lorenz--Mie--Debye formulation
for scattering by a sphere \c{Grimes1}, but do not lead to 
significantly different results for electrically small
spheres. In the limit $R_a\to 0$, expressions \r{eqpol2} reduce to 
\r{eqpol1} because
\begin{equation}
\lim_{R_a\to 0}\tau_{a/b} = 0\,.
\end{equation}

Clearly, the incorporation of particle size--dependence {\em via} 
\r{eqpol2} in \r{eqBrugg} leads to a coupling of
the relative permittivities and the relative permeabilities.

\section{Results and Conclusion}\label{rescon}

In order to investigate the properties of \r{eqBrugg}, let us 
simplify it for a two--constituent composite material: $L=2$.
Expressions  \r{eqBrugg} for the size--dependent Bruggeman approach 
then read as follows:
\begin{equation}
\left.\begin{array}{ll}
0=f_1\frac{\eps_1-\eps_{Br}}{\eps_1 
(1-2\tau_{1/Br})+2\eps_{Br}(1+\tau_{1/Br})}\\[5pt]
\quad +\,(1-f_1)\frac{\eps_2-\eps_{Br}}{\eps_2 
(1-2\tau_{2/Br})+2\eps_{Br}(1+\tau_{2/Br})}
\\[8pt]
0=f_1\frac{\mu_1-\mu_{Br}}{\eps_1 
(1-2\tau_{1/Br})+2\mu_{Br}(1+\tau_{1/Br})}\\[5pt]
\quad +\,(1-f_1)\frac{\mu_2-\mu_{Br}}{\mu_2 
(1-2\tau_{2/Br})+2\mu_{Br}(1+\tau_{2/Br})}
\end{array}\right\}\,.
\label{eqBr}
\end{equation}
These two coupled equations have to be solved together in order to 
obtain the estimates $\eps_{Br}$ and
$\mu_{Br}$ of $\eps_{HCM}$ and $\mu_{HCM}$ as functions of $k_0$, 
$f_1$, $R_1$ and $R_2$.

Equations  \r{eqBr} have to be solved iteratively, and the 
Newton--Raphson method is very useful
for that purpose \cite[Sec. 6.5.2]{CC}. Typically, this method 
requires an initial guess, which can be supplied
using the Maxwell Garnett approach \c{LOCM,Neela}. If $f_1> f_2$, 
then material $1$ should be treated as the host material while 
material $2$ is dispersed
in particulate form; and the size--dependent Maxwell Garnett 
estimates of $\eps_{HCM}$ and $\mu_{HCM}$ are then obtained as 
follows:
\begin{equation}
\left.\begin{array}{ll}
\eps_{MG,1} = \eps_1 + (1-f_1) 
\frac{\alpha_e^{2/1}}{1-(1-f_1)\frac{\alpha_e^{2/1}}{3\eps_1}}\\[5pt]
\mu_{MG,1} = \mu_1 + (1-f_1) 
\frac{\alpha_h^{2/1}}{1-(1-f_1)\frac{\alpha_h^{2/1}}{3\mu_1}}
\end{array}\right\}\,.
\label{eqMG1}
\end{equation}
On the other hand, the size--dependent Maxwell Garnett estimates
\begin{equation}
\left.\begin{array}{ll}
\eps_{MG,2} = \eps_2 + f_1 
\frac{\alpha_e^{1/2}}{1-f_1\frac{\alpha_e^{1/2}}{3\eps_2}}\\[5pt]
\mu_{MG,2} = \mu_2 + f_1 
\frac{\alpha_h^{1/2}}{1-f_1\frac{\alpha_h^{1/2}}{3\mu_2}}
\end{array}\right\}\,
\label{eqMG2}
\end{equation}
appear more appropriate when $f_2>f_1$. Incorporation of size 
dependence couples dielectric
and magnetic properties also in \r{eqMG1} and \r{eqMG2}.

\begin{figure}[!t]
\centering \psfull
\epsfig{file=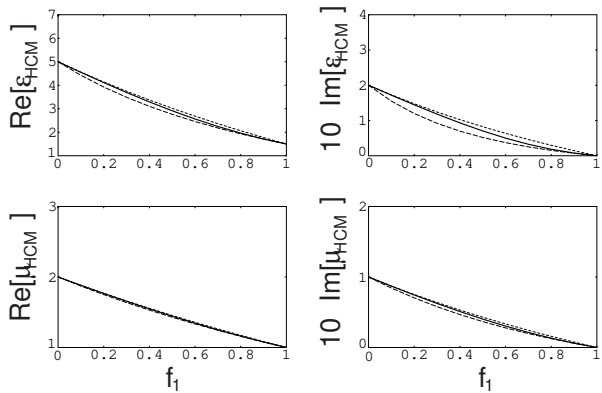,width=8cm }
\caption{Estimates of the real and imaginary parts of the relative 
permittivity and the relative permeability
of a homogenized composite material (HCM) with two constituent 
materials ($\eps_1=1.5$, $\mu_1=1$, $\eps_2=5+i0.2$,
and $\mu_2=2+i0.1$) as functions of the volumetric fraction
$f_1=1-f_2$. Size--independent Maxwell Garnett approach with material 
$1$ as the host material (dashed line);
Size--independent Maxwell Garnett approach with material $2$ as the 
host material (dotted line);
Size--independent Bruggeman approach   (solid line). $k_0R_1=k_0R_2=0$.
}
\label{Fig1}
\end{figure}

\begin{figure}[!h]
\centering \psfull
\epsfig{file=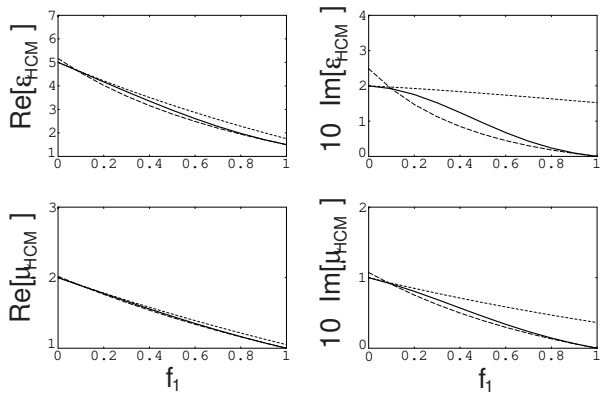,width=8cm }
\caption{Estimates of the real and imaginary parts of the relative 
permittivity and the relative permeability
of a homogenized composite material (HCM) with two constituent 
materials ($\eps_1=1.5$, $\mu_1=1$, $\eps_2=5+i0.2$,
and $\mu_2=2+i0.1$) as functions of the volume fraction
$f_1=1-f_2$. Size--dependent Maxwell Garnett approach with material 
$1$ as the host material (dashed line);
Size--dependent Maxwell Garnett approach with material $2$ as the 
host material (dotted line);
Size--dependent Bruggeman approach   (solid line). $k_0R_1=k_0R_2=0.2$.
}
\label{Fig2}
\end{figure}

\begin{figure}[!ht]
\centering \psfull
\epsfig{file=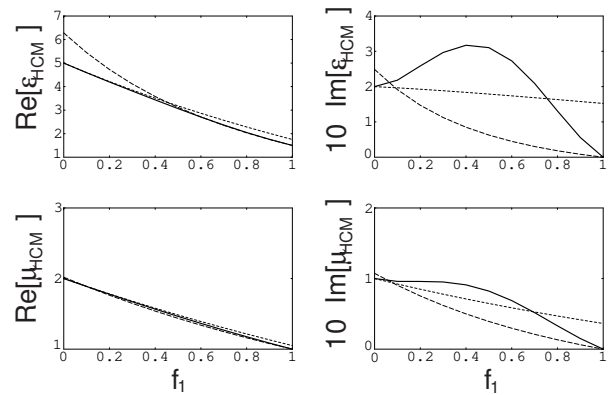,width=8cm }
\caption{Same as Figure \ref{Fig2}, except that $k_0R_1=0.2$ and $k_0R_2=0.6$.
}
\label{Fig3}
\end{figure}

Let us note that the limits
\begin{equation}
\left.\begin{array}{ll}
\lim_{f_1\to0} 
\left[\begin{array}{l}\eps_{Br}\\\mu_{Br}\end{array}\right] 
=\left[\begin{array}{l}\eps_{2}\\\mu_{2}\end{array}\right] \\[7pt]
\lim_{f_1\to1} 
\left[\begin{array}{l}\eps_{Br}\\\mu_{Br}\end{array}\right] 
=\left[\begin{array}{l}\eps_{1}\\\mu_{1}\end{array}\right]
\end{array}\right\}\,
\end{equation}
satisfied by the solutions of \r{eqBr}
are physically correct, and
are not affected by the incorporation of size dependence in the 
Bruggeman approach.
In contrast, the size--dependent Maxwell Garnett expressions 
\r{eqMG1} and \r{eqMG2} do not exhibit physically reasonable
limits when the host material vanishes; i.e.,
\begin{equation}
\left.\begin{array}{ll}
\lim_{f_1\to0} \left[\begin{array}{l}\eps_{MG,1}\\\mu_{MG,1}\end{array}\right]
\ne\left[\begin{array}{l}\eps_{2}\\\mu_{2}\end{array}\right]\,, 
&\quad{\rm if}\, R_2\ne 0 \\[7pt]
\lim_{f_1\to0} \left[\begin{array}{l}\eps_{MG,1}\\\mu_{MG,1}\end{array}\right]
=\left[\begin{array}{l}\eps_{2}\\\mu_{2}\end{array}\right]\,, 
&\quad{\rm if}\, R_2= 0 \\[7pt]
\lim_{f_1\to1} 
\left[\begin{array}{l}\eps_{MG,1}\\\mu_{MG,1}\end{array}\right] 
=\left[\begin{array}{l}\eps_{1}\\\mu_{1}\end{array}\right]
\end{array}\right\}\,,
\end{equation}
and analogously for $\eps_{MG,2}$ and $\mu_{MG,2}$.
The foregoing limits are borne out by the plots of $\eps_{HCM}$ and 
$\mu_{HCM}$  {\em versus\/} $f_1$ presented
in Figures \ref{Fig1}--\ref{Fig3}.

Figure \ref{Fig1} presents estimates of the real and imaginary parts 
of the relative permittivity and the relative permeability
of the HCM when $\eps_1=1.5$, $\mu_1=1$, $\eps_2=5+i0.2$,
and $\mu_2=2+i0.1$, and the sizes $R_1=R_2=0$. Calculations for the 
relative permittivity and the relative permeability
then decouple from each other.

The analogous plots in Figure \ref{Fig2} were drawn for $R_1=R_2\ne 
0$. These plots are quite different from those
in the preceding figure. The imaginary parts of $\eps_{HCM}$ and 
$\mu_{HCM}$ appear to be more affected
by the size dependence than the real parts. Indeed, were both 
constituent materials totally nondissipative, the imaginary
parts of $\tau_{a/b}$--terms would still give rise to imaginary parts 
of both $\eps_{HCM}$ and $\mu_{HCM}$ \c{PLB}.
We also conclude from comparing Figures \ref{Fig1} and \ref{Fig2}
that dielectric--magnetic coupling proportionally affects the 
imaginary parts of the HCM constitutive parameters
more than their real parts.

There is no reason for the particles of both constituent materials to 
be of the same size (or have the same
distribution of size, in general). The plots in Figure \ref{Fig3} 
were drawn for $R_2=3R_1$. Clearly from this figure and
Figure \ref{Fig2}, the effect
of different particle sizes on dielectric--magnetic coupling can be 
substantial.

The permeability contrast between the two constituent materials chosen
for Figures \ref{Fig1}--\ref{Fig3} is less than the permittivity 
contrast. We notice that the effect of
size dependence on $\mu_{HCM}$ is less than on $\eps_{HCM}$. This 
implies that dielectric--magnetic
coupling affects the more contrasting constitutive parameter more.

To conclude, we have here implemented the Bruggeman approach for the 
homogenization of
dielectric--magnetic composite materials, without ignoring the sizes
of the
spherical particles. These expressions exhibit the proper limit 
behavior. The incorporation
of size dependence is directly responsible for the emergence of 
dielectric--magnetic coupling in the
estimated relative permittivity and permeability of the homogenized 
composite material. The size--dependent Bruggeman
estimates are compared with the size--dependent Maxwell Garnet 
estimates, which do not necessarily evince the proper limit
behavior and are therefore applicable to dilute composite materials.








\Biography{Akhlesh Lakhtakia}{ was born in Luck\-now, India, in
1957. Presently, he is a Distinguished Pro\-fessor of Engineering Science and
Mech\-anics
at the Penn\-sylania State Uni\-versity. He is a Fellow of the
Optical Society of America, SPIE--The International Society for Optical
Engineering, and the Institute of Physics (United Kingdom). He has
either author\-ed
or co--author\-ed about 650
journal papers and conference publications, and has lectured on
waves and complex mediums in many countries.
His current research interests lie in the
electro\-magnetics  of complex mediums,
sculpt\-ured
thin films, and nano\-technology. For more information
on his activities, please visit
his website: www.esm.psu.edu/$\sim$axl4/}

\Biography{Tom G. Mackay}{
graduated MSci in Mathematics from
the University of Glasgow, UK,  in 1998, after spending the previous ten
years working as a bioengineer at Glasgow Royal Infirmary. He
spent the next three years  engaged in postgraduate studies in the
Department of Mathematics at the University of Glasgow, under the
supervision of Prof. Werner S. Weiglhofer. Upon completing his PhD
thesis in 2001, he moved to the University of Edinburgh, UK,  where he
is currently employed as a lecturer in the School of Mathematics.
His research interests are primarily related to the homogenization
of complex electromagnetic systems. He is also interested in biological
applications
of electromagnetic theory.}

\end{document}